\documentclass[aip,jap,preprint]{revtex4-1}

\usepackage{color}
\usepackage{soul}
\usepackage{setspace}
\usepackage{gensymb}
\usepackage{graphicx}
\usepackage{epstopdf}
\usepackage{amssymb}
\usepackage[colorlinks=true,linkcolor=blue]{hyperref}
\usepackage{array}
\usepackage{bm}
\usepackage{float}
\usepackage{amsmath}
\usepackage{verbatim}
\graphicspath{ {figures/} }
\usepackage{listings}
\usepackage{mathtools}

\begin{document}
	
	\title{MaGPoS - A novel decentralized consensus mechanism combining magnetism and proof of stake}
	
	\date{\today}
	
	\author{Tommy~Mckinnon - admin@hyperevo.com}

	\begin{abstract}
	\noindent We describe MaGPoS, a novel consensus mechanism which is well suited to decentralized blockchain based protocols. MaGPoS is based on a combination of the well known physics of nano-scale magnetism, and previous implementations of proof of stake. This system has been studied by hundreds of thousands of scientists worldwide for over a hundred years, giving it an extreme level of reliability that is needed for a consensus mechanism. We start by explaining the physics, and study the properties that make it particularly beneficial for use in a consensus mechanism. We then show how to apply the physical model to a decentralized network of nodes, each with their own copy of a blockchain. After this, we describe some example calculations that a node in the decentralized network would make, and provide pseudo code for implementation. Finally, we discuss the how the model achieves all of the important properties that one expects of a consensus mechanism.

	\end{abstract}

	\maketitle
	
	For years, many different cryptocurrencies all shared the same consensus mechanism as described in the original bitcoin whitepaper, called proof of work (PoW). PoW is a very elegant solution to the problem of achieving consensus in a decentralized network because it uses the limited resource of computational power to ensure that no one entity can gain control over the network. It has worked so well that the term blockchain is now synonymous with immutability and reliability. However, as blockchain has grown in popularity, a massive problem has arisen. Miners all over the world are using all of the computing resources they have to compete with one another, which has resulted in a massive amount of wasted energy and resources. What makes it worse is that the computations that are being done have no value whatsoever. Bitcoin, one of the most popular cryptocurrencies today, uses the same amount of energy as the entire country of the Czech Republic. \cite{cite_4} Considering that $38.4$\% and $23.2$\% of the power generation globally comes from coal and natural gas, respectively, that is a massive amount of CO$_2$ being pumped into the atmosphere. \cite{world_key_energy_stat} With the imminent threat of climate change, which is driven by CO$_2$ production, this is a serious threat to our planet and the future of humanity.
	
	Recently, there has been a big push towards a new type of consensus mechanism, that solves the energy usage problems associated with PoW, known as proof of stake (PoS). Instead of using computational power as the limited resource to achieve consensus, PoS uses capital, also known as stake. There have been several different forms of PoS, with each new iteration improving on the one before it, most notably, the original PeerCoin consensus mechanism, Ethereum Casper, and PoSV2. \cite{king2017ppcoin, casper_1, casper_2, vasin2014blackcoin} 
	
	In this paper, we will present a new iteration of PoS, which we have called MaGPoS. This new consensus mechanism is based on the quantum mechanics of a lattice of magnetic dipoles in a nano-scale thin film. In such a system, with the absence of external magnetic fields, anisotropies, and demagnetization fields, the exchange coupling between the dipoles always results in them aligning themselves after being disturbed. In order to utilize this phenomenon for our consensus mechanism, we can think of each dipole as a node in the decentralized network, and the direction at which the dipole points is analogous to the blockchain fork choice of that node. If two nodes have chosen the same fork, they are pointing in the same direction, and if they have chosen a different fork, they are pointing in a different direction. Finally, by applying the physics of exchange coupling between the nodes, we can use physics to prove that they will come to consensus with one another and agree on a single fork. We can also use physics to prove that the choice of the fork will be that which the stake-weighted majority of the nodes on the network have chosen. This is an important property of any consensus mechanism used in financial systems because it gives predictable and reliable results every time. 
	
	This approach is superior to other consensus mechanisms for several reasons that will be explained throughout this paper. The biggest advantages are:
	
	\begin{enumerate}
		\item Mathematical provability and predictability, with reliability that is backed by physics.
		\item No need for a separate group of validators. In MaGPoS, the nodes are validators.
		\item Highly scalable. Nodes only need to communicate with the closest neighbors within the XOR topology of the network. This saves on time and bandwidth, neither of which increase as the network scales and grows.
		\item Extremely low energy usage.
		\item Full decentralization.
		\item Strong immutability.
		\item Highly resistant to forks.
	\end{enumerate}
	
	\section{Brief overview of the physics}
	
	The direction of magnetization of a dipole is determined by the minimum of the total free-energy, which can be expressed as the sum:
	
	\begin{equation}\label{eq:energy_1}
	E_{\rm{tot}} = E_{\rm{ex}} + E_{\rm{Zeeman}} + E_{\rm{ani}} + E_{\rm{demag}},
	\end{equation}
	where $E_{\rm{ex}}$ is the energy associated with the exchange interaction,  $E_{\rm{Zeeman}}$ is the energy associated with any external magnetic fields, $E_{\rm{ani}}$ is the energy associated with the crystallographic or interface anisotropy, $E_{\rm{demag}}$ is the energy associated with the demagnetization field caused by the minimization of magnetic charge on the surface of the sample. \cite{montoya2016spin,doi:10.1063/1.5045697,tommy1}
	
	In ultrathin polycrystalline magnetic layers, both the demag and anisotropy contributions are both in the same uniaxial direction. In some layer structures, such as Ta/FeCoB/MgO, the thickness of the FeCoB layer can be chosen such that the anisotropy arising from the FeCoB/MgO interface cancels out the demag contribution. \cite{doi:10.1063/1.4746426} In this case, the anisotropy and demag contribution have an equal and opposite effect, allowing them to be ignored. Furthermore, the sample can be placed in an area with no external magnetic fields, removing the Zeeman contribution. With these modifications, the total free energy becomes:
	
	\begin{equation}\label{eq:energy_2}
	E_{\rm{tot}} = E_{\rm{ex}}.
	\end{equation}
	
	Using the Heisenberg model for a 2 dimensional lattice of magnetic dipoles, which are created by spins of unpaired electrons, the exchange energy can be written as: \cite{getzlaff2007fundamentals}

	\begin{equation}\label{eq:energy_3}
	E_{\rm{tot}} = - \sum_{i,j} J_{i,j} \bm{S}_i \cdot \bm{S}_j,
	\end{equation}
	where $J_{i,j}$ is the coupling constant between electron spins $i$ and $j$, which is determined by properties of the material, $\bm{S}_i$ and $\bm{S}_j$ are the wave functions of the quantized spins $i$ and $j$, respectively (+1/2 or -1/2). 
	
	The total energy of the magnetic material is found by summing over all spins $i$ and $j$. However, because the exchange interaction drops off so rapidly with distance, only summation over nearest neighbor spins is necessary. \cite{getzlaff2007fundamentals} Just like all physical systems, this one will choose the direction of all spins such that the free energy is minimized. In this case, it is easy to see that it occurs when all of the spins are pointing in the same direction.
	
\begin{figure}[h]
		\includegraphics[width=70mm,scale=1]{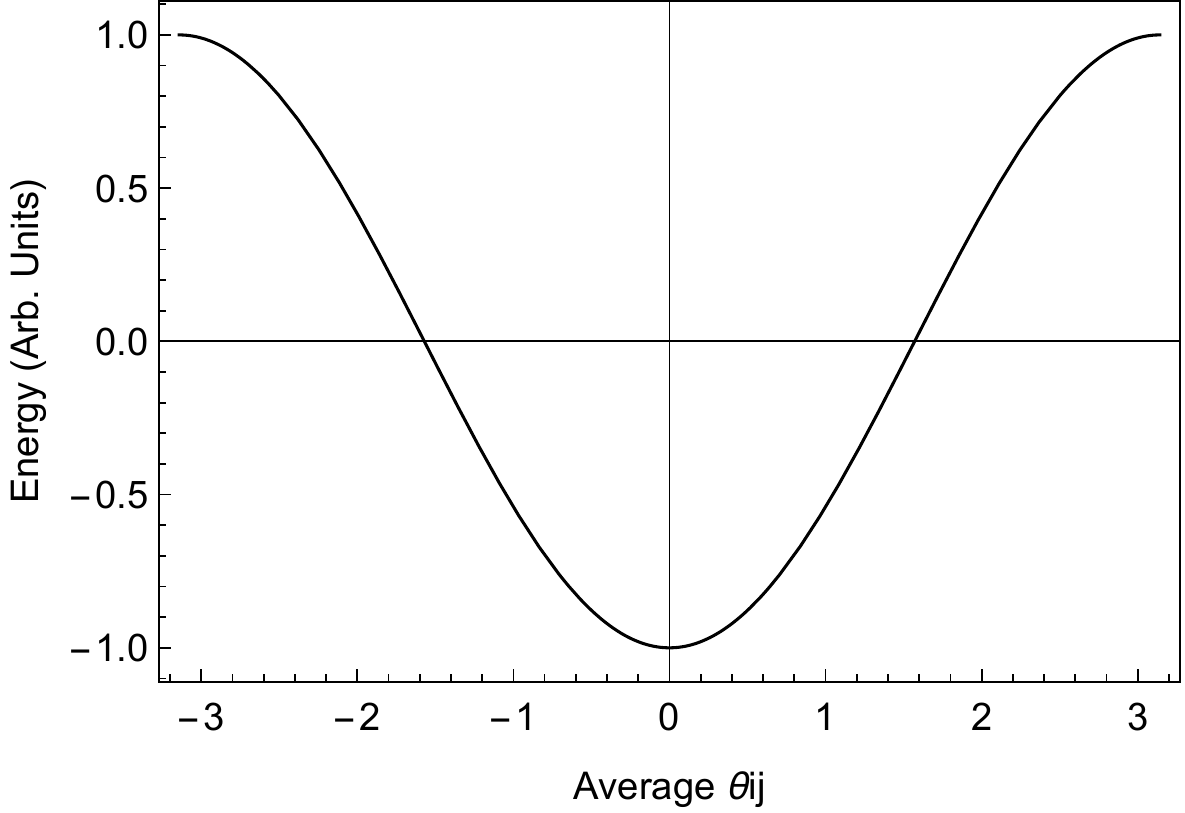}
		\caption{Free energy of the system as a function of the average angle between the magnetic moments of all of the dipoles. The energy minima occurs when the angle is 0 and all of the dipoles are aligned.}
		\label{fig:111}
	\end{figure}

	In order to illustrate this, lets look at the classical case where we have a very large number of spins. We replace the spin dot product in Eq. \ref{eq:energy_3} with a dot product of the magnetic moments of dipoles $i$ and $j$ in the lattice. In this case, the free energy would look like
	
	\begin{equation}\label{eq:energy_4}
	E_{\rm{tot}} = - \sum_{i,j} J_{i,j} \cos(\theta_{i,j}) M_i M_j,
	\end{equation}
	where we have now written out the dot product explicitly and $\theta_{i,j}$ is the angle between the magnetic moment of dipoles $i$, and $j$. The free energy as a function of the average angle of the magnetic moments across all of the dipoles can be seen in Fig. \ref{fig:111}. The energy minima occurs when the angle is 0 and all of the dipoles are aligned. Thus, the system will always come to equilibrium with the dipoles being aligned.
	
\begin{figure}[h]
		\includegraphics[width=70mm,scale=1]{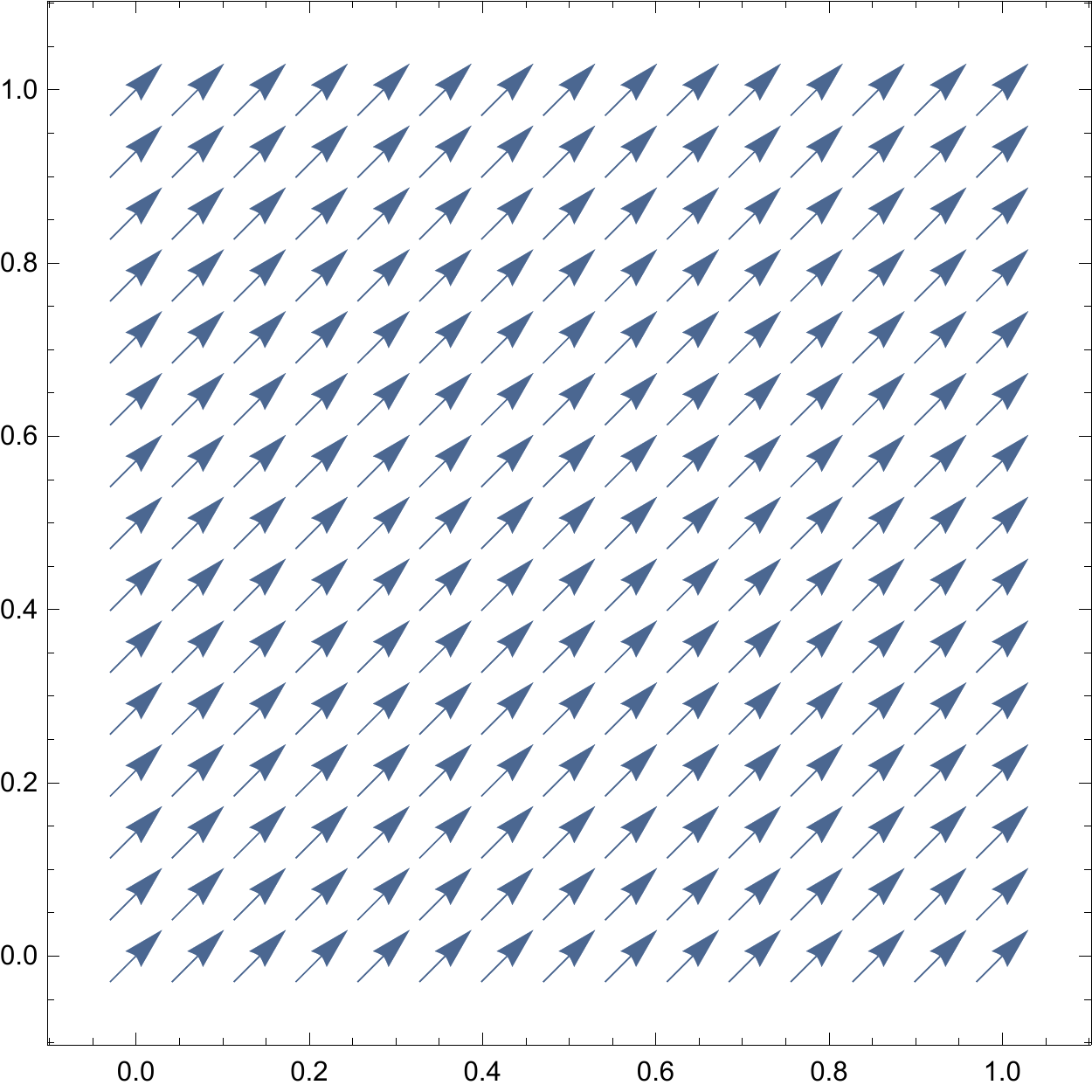}
		\caption{An grid of magnetic dipoles, where each arrow is a dipole. All of the dipoles are oriented parallel to each other due to the exchange interaction between them. Each magnetic dipole is analogous to a node in a decentralized network. The parallel directions of the magnetic dipoles are analogous to nodes being in agreement on their fork choice, and being in consensus.}
		\label{fig:3}
	\end{figure}

	We can visualize what a two dimensional grid of dipoles, which are in equilibrium at the energy minima, would look like, which has been shown in Fig \ref{fig:3}. 
	
	\begin{figure*}[t]
		\includegraphics[width=160mm]{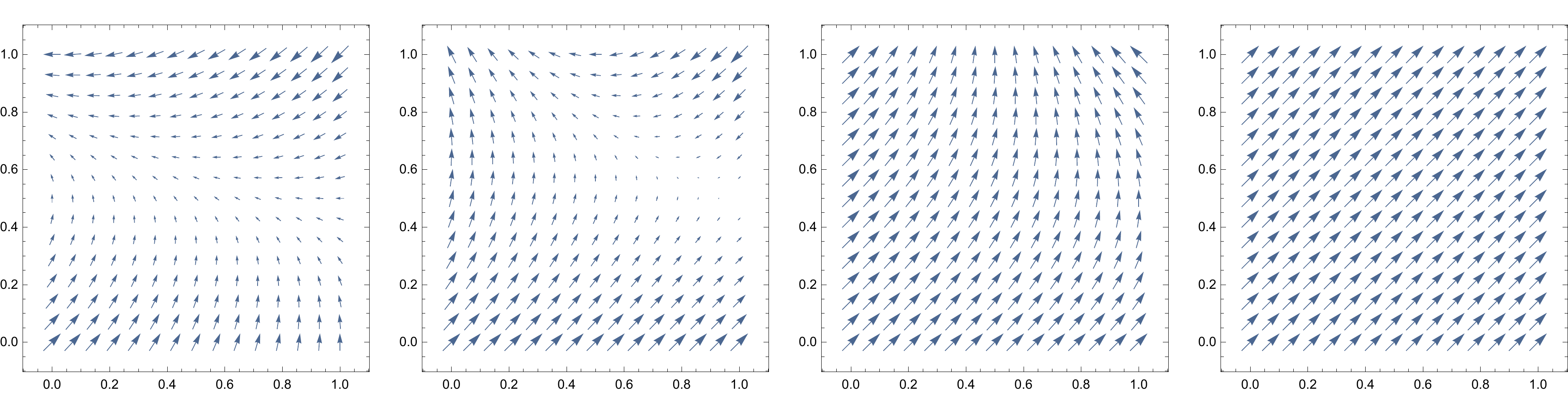}
		\caption{From left to right the exchange interaction causes the magnetic dipoles to align parallel with each other. This is analogous to the nodes on the network reaching consensus.}
		\label{fig:4}
	\end{figure*}

	Now we can imagine forcing the magnetic dipoles at the bottom left and top right corners of the lattice to point in opposite directions as shown in Fig.~\ref{fig:4}. This will cause a magnetic domain wall to appear between the bottom left and top right dipoles. If we then let go of the magnetic dipoles at the corners, and let the system reach equilibrium, the free energy between the magnetic dipoles will cause them all to align again. This process is shown in from left to right in Fig.~\ref{fig:4}.
	
	The important thing to understand here, is if any magnetic dipoles are pointing in different directions, the interactions between them will cause them to realign themselves to point in the same direction.

	\section{Application to the consensus mechanism}
	Now that we have a basic understanding of the physics, and how a lattice of dipoles spontaneously aligns themselves, we will show how we apply it to create a consensus mechanism with the same properties. The first thing that we do is define each node on the decentralized network as a single dipole moment in the two dimensional lattice of dipoles. In this analogy, the distance between the nodes is not given by the physical distance between then, instead, it is given by the XOR distance between the nodes in the Kademlia distributed hash table. 
	
	Just like the dipoles, each node in the network has a free energy that can be described with Eq. \ref{eq:energy_3}, except instead of $\bm{S}_i$ referring to the spin of an electron in the lattice, it now refers to the fork choice. $\bm{S}_i$ is now an n-dimensional vector, where n is the total number of different forks available, and each fork points in a direction orthogonal to all other forks. In this case, $\bm{S}_i \cdot \bm{S}_j = 0$ if the two forks are different, and $\bm{S}_i \cdot \bm{S}_j = 1$ if they are the same. Another critical difference is that the coupling constant $J_{i,j}$ is now equal to the stake of the remote node, which we have indexed as $j$. This is where the concept of PoS comes into this consensus mechanism. Weighting each node's contribution by their respective stake is required for all of the same reasons as in standard PoS. The biggest of which is to stop someone from setting up a large number of nodes and increasing their vote weight.
	
	Now, just like with the magnetic system, the nodes have one goal: to minimize their free energy. In order for the nodes to do this, they ask all closest neighbor nodes for their fork choices. Note, this only occurs when a conflict is found so that bandwidth is conserved. The nodes will then receive all of the fork choices, and go through all of the possible scenarios for it's local fork choice, and choose the one that results in the minimum energy.
	
	\subsection{Example calculation}
	
	Lets assume there are 5 nodes in our test system. The node stake and fork choices are shown here:
	\begin{center}
		\begin{tabular}{ c | c | c }
			Node ID & Stake & Fork Choice \\ 
			\hline
			0 & 10 & x \\  
			1 & 50 & x \\ 
			2 & 20 & y \\
			3 & 30 & z \\
			4 & 80 & z \\
		\end{tabular}
	\end{center}

	Lets pretend we are node 1, and we are trying to calculate the correct fork choice. The calculation we want to do, is cycle through all possible local fork choices, calculate the energy in each case, and go with the one that results in the minimum energy. From the point of view of a single node, and only looking at the contribution of nearest neighbor nodes, the energy equation simplifies to:
	
	\begin{equation}\label{eq:energy_33}
	E_{\rm{tot}} = - \sum_{i} J_{i} \bm{S}_0 \cdot \bm{S}_i,
	\end{equation}
	where the node calculating the energy is the node with index 0.
	We will first calculate the free energy if the local fork choice was x:
	
	\begin{equation}\label{eq:energy_5}
		\begin{aligned}
		E_{\rm{tot,x}}& = -1 \cdot (J_{0,0} \bm{S}_0 \cdot \bm{S}_0 + J_{0,1} \bm{S}_0 \cdot \bm{S}_1 + \\
		& J_{0,2} \bm{S}_0 \cdot \bm{S}_2 + J_{0,3} \bm{S}_0 \cdot \bm{S}_3 + \\
		& J_{0,4} \bm{S}_0 \cdot \bm{S}_4)\\
		& = -1 \cdot (10 + 50 - 20 -30 - 80)\\
		& = 70
		\end{aligned}
	\end{equation}
	
	Note, we make sure to calculate the self energy where $i$ = $j$ = 0, this ensures that we count the contribution from our local fork choice and stake. Next, we will calculate the free energy if the local fork choice was y:
	
	\begin{equation}\label{eq:energy_6}
		\begin{aligned}
		E_{\rm{tot,y}}& = -1 \cdot (-10 - 50 + 20 -30 - 80)\\
		& = 150
		\end{aligned}
	\end{equation}
	
	Next, we will calculate the free energy if the local fork choice was z:
	
	\begin{equation}\label{eq:energy_7}
		\begin{aligned}
		E_{\rm{tot,z}}& = -1 \cdot (-10 - 50 - 20 + 30 + 80)\\
		& = -30
		\end{aligned}
	\end{equation}
	
	We can see that fork choice z resulted in the minimum free energy, therefore we choose that fork and say fork z is in consensus. This will cause node 0 to change from fork x to fork z, further lowering the energy of fork z, which thereby causes other nodes around to also choose for z. This mechanism propagates a single dominant fork choice across the network. In such a simple network, this calculation closely approximates taking a kind of stake weighted average fork choice. However, the real power of this model comes in when the network becomes very large. In this model, each node only needs information from the nearest neighbor nodes, it never needs to communicate any long range distance in the network. In other PoS models, such as Ethereum Casper, each node needs to accumulate votes from a large fraction of the validators before it knows what fork is correct. This takes more time, and uses much more bandwidth. Furthermore, the time and bandwidth required for our model is constant and doesn't depend on the size of the network, making it a perfect solution for a scalable blockchain. 
	
	\clearpage
	\subsection{Pseudo code}	
	\begin{lstlisting}
if conflict found:
	ask all connected peers for their fork choice
	
	lowest_energy = infinity
	fork_choice_with_lowest_energy = None
	
	for fork_choice in all_fork_choices:
		energy = 0
		for peer in (connected_peers + this_node):
			if fork_choice = peer.fork_choice:
				energy -= peer.stake
			else:
				energy += peer.stake
		if energy < lowest_energy:
			lowest_energy = energy
			fork_choice_with_lowest_energy = fork_choice
	
	if this_node.fork_choice is not fork_choice:
		change this node's fork choice to the new fork choice
	\end{lstlisting}
	
	\section{Discussion}
	When choosing a consensus mechanism for a blockchain project, there are a few aspects that are absolutely essential:
	
	\begin{enumerate}
		\item The consensus mechanism always results in exactly one choice that the entire network agrees on. Or in other words, no spontaneous forks.
		\item Immutability: We can reliably assume blocks won't be changing.
		\item Decentralization: Make sure no one entity has more control than others. This is an area where delegated PoS fails.
		\item Scalability: Projects should always be prepared for growth in the future as mass adoption takes hold.
	\end{enumerate}
	
	Because MaGPoS is basically a simulated physics environment that mirrors the rules of nature, we can ensure that the outcome will also be the same as is observed in nature, and is well understood using mathematics. This immediately satisfies 1. because, as discussed above, requiring that energy is always minimized guarantees that all of the nodes will come to the same fork choice.
	
	How MaGPoS achieves immutability might be a bit less obvious, but it does have the same level immutability as any other PoS implementation. If you look at Fig. \ref{fig:3}, you can imagine grabbing one of the arrows and changing it's direction. This is equivalent to a rogue node that is proposing a new fork that none of the other nodes have, likely in an attempt to remove some transaction that already took place in the competing fork. In this case, there is a much larger number of nodes with the other fork, and they will force the less popular fork out of the network. What happens, is all of the nodes calculate their own minimum energy, and this will always result in them choosing the fork that has the majority of the network's stake. We will now prove this mathematically. Lets take the simple case where there are just 2 competing forks. One fork has the majority of the stake, and the second has the minority of the stake. The free energy, from the point of view of a single node, can be written as:
	
	\begin{equation}\label{eq:energy_10}
	\begin{aligned}
	E_{\rm{tot}} & = - \sum_{i = 0}^{N} J_{i} \bm{S}_0 \cdot \bm{S}_i \\
	& \underbrace{- \sum_{i = 0}^{n < N/2} J_{i} \bm{S}_0 \cdot \bm{S}_i}_\text{\clap{minority}} - \underbrace{\sum_{i = n}^{N} J_{i} \bm{S}_0 \cdot \bm{S}_i}_\text{\clap{majority}},
	\end{aligned}
	\end{equation}
	where we have separated the contribution from the minority fork and majority fork. If we now go through the process of calculating which fork results in the lowest energy, we have two cases to try. The first case, is if we go with the fork choice of the minority, which gives the energy:
	
	\begin{equation}\label{eq:energy_11}
	\begin{aligned}
	E_{\rm{tot,min}}& = - \sum_{i = 0}^{n < N/2} J_{i} \cdot 1 - \sum_{i = n}^{N} J_{i} \cdot (-1) \\
	& = \sum_{i = n}^{N} J_{i} - \sum_{i = 0}^{n < N/2} J_{i}. \\
	\end{aligned}
	\end{equation}
	
	Next, we calculate the energy of the case where this node has the majority stake:
	\begin{equation}\label{eq:energy_12}
	\begin{aligned}
	E_{\rm{tot,maj}}& = - \sum_{i = 0}^{n < N/2} J_{i} \cdot (-1) - \sum_{i = n}^{N} J_{i} \cdot 1 \\
	& = \sum_{i = 0}^{n < N/2} J_{i} - \sum_{i = n}^{N} J_{i}. \\
	\end{aligned}
	\end{equation}
	
	Next, we subtract the minority energy from the majority energy
	\begin{equation}\label{eq:energy_13}
	\begin{aligned}
	E_{\rm{tot,min}} - E_{\rm{tot,maj}}& = \frac{1}{2}\bigg[\sum_{i = n}^{N} J_{i} - \sum_{i = 0}^{n < N/2} J_{i}\bigg] > 0
	\end{aligned}
	\end{equation}
	
	Now, as long as this always results in a number that is greater than 0, then the energy when going with the minority case will be greater than the energy when going with the majority case, which will result in the majority case being chosen. Looking at this equation, it is easy to see that all we are doing is adding up the total stake of all the nodes with the majority fork, and subtracting the total stake of all of the nodes with the minority fork. This will be positive as long as the nodes in the majority case also contain the majority of the stake. Therefore, there could exist a situation where the minority fork choice ends up being chosen because it has more stake. This is by design, because an attacker could spin up a large number of nodes with their rogue fork, and easily become the majority of the nodes, but the network would still not care because of the lack of stake. Thus, the MaGPoS consensus mechanism will always end up choosing the fork that the nodes with the majority of the stake have chosen. 
	
	If you were following along closely, you will have realized that in the pseudo code, the nodes only need to ask the closest neighbor nodes, but in this calculation we did a sum over all nodes. However, this is not a problem because it has been shown elsewhere that physically they give equivalent results. \cite{getzlaff2007fundamentals}
	
	This result means that it would be very difficult to cause the network to switch forks, or revert a block, after it has been distributed to the network and reached consensus. The only possible way for this to occur is if the attacker had more than 50\% of the total stake of the network, also known as a 51\% attack, which is a vulnerability of all PoS consensus mechanisms.
	
	MaGPoS is fully decentralized because of two reasons. The first is that every single node participates in deciding which fork the network will choose. Every node also gets to contribute a vote that is weighted by their stake. There is no authority or central power than can tell the network which fork to choose. The second aspect that makes this consensus mechanism particularly decentralized is the fact that nodes only need to communicate with it's local peers. So nodes on one side of the XOR space are completely disconnected from nodes on the other side. They are working completely independently, which is a form of decentralization.
	
	Like other PoS implementations, MaGPoS uses thousands of times less energy than PoW. This naturally makes it highly scalable because the network can grow in usage and popularity without wasting any resources or contributing unnecessarily to climate change. Not only that, MaGPoS also has a property that makes it more scalable than some other PoS alternatives. Nodes only need to communicate with nearest neighbor nodes. This number doesn't increase as the network scales and more nodes are added to the XOR space. Thus, the network bandwidth, and time required to calculate consensus doesn't increase as the network scales. This is unlike some other implementations that require accumulating votes from a large number of validators before knowing the true fork choice.
	
	\subsection{Slashing}
	Like Ethereum Casper, MaGPoS can have slashing enabled. Slashing is a further way of penalizing nodes that break the rules, and it increases the immutability. The implementation of slashing on MaGPoS would be virtually identical to Ethereum Casper, so we will simply refer you to their whitepaper at Ref. \cite{casper_1}.
	
	\section{Real world example}
	We are implementing MaGPoS into our upcoming fully open source project called Helios Protocol \\
	https://heliosprotocol.io/. We have been developing it for over a year now, and plan on releasing mainnet in multiple phases over the second half of 2019. This will be the first real-world test of the MaGPoS consensus mechanism, and will allow us to prove the concept.

	\bibliography{mybib}

\begin{thebibliography}{11}%
\makeatletter
\providecommand \@ifxundefined [1]{%
 \@ifx{#1\undefined}
}%
\providecommand \@ifnum [1]{%
 \ifnum #1\expandafter \@firstoftwo
 \else \expandafter \@secondoftwo
 \fi
}%
\providecommand \@ifx [1]{%
 \ifx #1\expandafter \@firstoftwo
 \else \expandafter \@secondoftwo
 \fi
}%
\providecommand \natexlab [1]{#1}%
\providecommand \enquote  [1]{``#1''}%
\providecommand \bibnamefont  [1]{#1}%
\providecommand \bibfnamefont [1]{#1}%
\providecommand \citenamefont [1]{#1}%
\providecommand \href@noop [0]{\@secondoftwo}%
\providecommand \href [0]{\begingroup \@sanitize@url \@href}%
\providecommand \@href[1]{\@@startlink{#1}\@@href}%
\providecommand \@@href[1]{\endgroup#1\@@endlink}%
\providecommand \@sanitize@url [0]{\catcode `\\12\catcode `\$12\catcode
  `\&12\catcode `\#12\catcode `\^12\catcode `\_12\catcode `\%12\relax}%
\providecommand \@@startlink[1]{}%
\providecommand \@@endlink[0]{}%
\providecommand \url  [0]{\begingroup\@sanitize@url \@url }%
\providecommand \@url [1]{\endgroup\@href {#1}{\urlprefix }}%
\providecommand \urlprefix  [0]{URL }%
\providecommand \Eprint [0]{\href }%
\providecommand \doibase [0]{http://dx.doi.org/}%
\providecommand \selectlanguage [0]{\@gobble}%
\providecommand \bibinfo  [0]{\@secondoftwo}%
\providecommand \bibfield  [0]{\@secondoftwo}%
\providecommand \translation [1]{[#1]}%
\providecommand \BibitemOpen [0]{}%
\providecommand \bibitemStop [0]{}%
\providecommand \bibitemNoStop [0]{.\EOS\space}%
\providecommand \EOS [0]{\spacefactor3000\relax}%
\providecommand \BibitemShut  [1]{\csname bibitem#1\endcsname}%
\let\auto@bib@innerbib\@empty
\bibitem [{\citenamefont {Digiconomist}(2018)}]{cite_4}%
  \BibitemOpen
  \bibfield  {author} {\bibinfo {author} {\bibnamefont {Digiconomist}},\ }\href
  {https://digiconomist.net/bitcoin-energy-consumption} {\enquote {\bibinfo
  {title} {Bitcoin energy consumption index},}\ } (\bibinfo {year}
  {2018})\BibitemShut {NoStop}%
\bibitem [{\citenamefont {Birol}(2018)}]{world_key_energy_stat}%
  \BibitemOpen
  \bibfield  {author} {\bibinfo {author} {\bibfnamefont {F.}~\bibnamefont
  {Birol}},\ }\href
  {https://webstore.iea.org/download/direct/2291?fileName=Key_World_2018.pdf}
  {\emph {\bibinfo {title} {World Energy Statistics 2018}}}\ (\bibinfo
  {publisher} {International Energy Agency},\ \bibinfo {year}
  {2018})\BibitemShut {NoStop}%
\bibitem [{\citenamefont {King}\ and\ \citenamefont
  {Nadal}(2017)}]{king2017ppcoin}%
  \BibitemOpen
  \bibfield  {author} {\bibinfo {author} {\bibfnamefont {S.}~\bibnamefont
  {King}}\ and\ \bibinfo {author} {\bibfnamefont {S.}~\bibnamefont {Nadal}},\
  }\href@noop {} {\bibfield  {journal} {\bibinfo  {journal} {URL
  https://peercoin. net/assets/paper/peercoin-paper. pdf.[Online}\ } (\bibinfo
  {year} {2017})}\BibitemShut {NoStop}%
\bibitem [{\citenamefont {Zamfir}(2013)}]{casper_1}%
  \BibitemOpen
  \bibfield  {author} {\bibinfo {author} {\bibfnamefont {V.}~\bibnamefont
  {Zamfir}},\ }\href
  {https://github.com/ethereum/research/blob/master/papers/CasperTFG/CasperTFG.pdf}
  {\enquote {\bibinfo {title} {{Casper the Friendly Ghost A
  "Correct-by-Construction" Blockchain Consensus Protocol}},}\ } (\bibinfo
  {year} {2013})\BibitemShut {NoStop}%
\bibitem [{\citenamefont {Buterin}\ and\ \citenamefont
  {Griffith}(2017)}]{casper_2}%
  \BibitemOpen
  \bibfield  {author} {\bibinfo {author} {\bibfnamefont {V.}~\bibnamefont
  {Buterin}}\ and\ \bibinfo {author} {\bibfnamefont {V.}~\bibnamefont
  {Griffith}},\ }\href {http://arxiv.org/abs/1710.09437} {\bibfield  {journal}
  {\bibinfo  {journal} {CoRR}\ }\textbf {\bibinfo {volume} {abs/1710.09437}}
  (\bibinfo {year} {2017})},\ \Eprint {http://arxiv.org/abs/1710.09437}
  {arXiv:1710.09437} \BibitemShut {NoStop}%
\bibitem [{\citenamefont {Vasin}(2014)}]{vasin2014blackcoin}%
  \BibitemOpen
  \bibfield  {author} {\bibinfo {author} {\bibfnamefont {P.}~\bibnamefont
  {Vasin}},\ }\href@noop {} {\bibfield  {journal} {\bibinfo  {journal} {URL:
  https://blackcoin. co/blackcoin-pos-protocol-v2-whitepaper. pdf}\ }\textbf
  {\bibinfo {volume} {71}} (\bibinfo {year} {2014})}\BibitemShut {NoStop}%
\bibitem [{\citenamefont {Montoya}(2016)}]{montoya2016spin}%
  \BibitemOpen
  \bibfield  {author} {\bibinfo {author} {\bibfnamefont {E.}~\bibnamefont
  {Montoya}},\ }\emph {\bibinfo {title} {Spin pumping and spin transport in
  magnetic heterostructures}},\ \href@noop {} {Ph.D. thesis},\ \bibinfo
  {school} {Science: Department of Physics} (\bibinfo {year}
  {2016})\BibitemShut {NoStop}%
\bibitem [{\citenamefont {McKinnon}\ and\ \citenamefont
  {Girt}(2018)}]{doi:10.1063/1.5045697}%
  \BibitemOpen
  \bibfield  {author} {\bibinfo {author} {\bibfnamefont {T.}~\bibnamefont
  {McKinnon}}\ and\ \bibinfo {author} {\bibfnamefont {E.}~\bibnamefont
  {Girt}},\ }\href {\doibase 10.1063/1.5045697} {\bibfield  {journal} {\bibinfo
   {journal} {Applied Physics Letters}\ }\textbf {\bibinfo {volume} {113}},\
  \bibinfo {pages} {192407} (\bibinfo {year} {2018})},\ \Eprint
  {http://arxiv.org/abs/https://doi.org/10.1063/1.5045697}
  {https://doi.org/10.1063/1.5045697} \BibitemShut {NoStop}%
\bibitem [{\citenamefont {McKinnon}\ \emph {et~al.}(2018)\citenamefont
  {McKinnon}, \citenamefont {Omelchenko}, \citenamefont {Heinrich},\ and\
  \citenamefont {Girt}}]{tommy1}%
  \BibitemOpen
  \bibfield  {author} {\bibinfo {author} {\bibfnamefont {T.}~\bibnamefont
  {McKinnon}}, \bibinfo {author} {\bibfnamefont {P.}~\bibnamefont
  {Omelchenko}}, \bibinfo {author} {\bibfnamefont {B.}~\bibnamefont
  {Heinrich}}, \ and\ \bibinfo {author} {\bibfnamefont {E.}~\bibnamefont
  {Girt}},\ }\href {\doibase 10.1063/1.5024949} {\bibfield  {journal} {\bibinfo
   {journal} {Journal of Applied Physics}\ }\textbf {\bibinfo {volume} {123}},\
  \bibinfo {pages} {223903} (\bibinfo {year} {2018})},\ \Eprint
  {http://arxiv.org/abs/https://doi.org/10.1063/1.5024949}
  {https://doi.org/10.1063/1.5024949} \BibitemShut {NoStop}%
\bibitem [{\citenamefont {Sokalski}\ \emph {et~al.}(2012)\citenamefont
  {Sokalski}, \citenamefont {Moneck}, \citenamefont {Yang},\ and\ \citenamefont
  {Zhu}}]{doi:10.1063/1.4746426}%
  \BibitemOpen
  \bibfield  {author} {\bibinfo {author} {\bibfnamefont {V.}~\bibnamefont
  {Sokalski}}, \bibinfo {author} {\bibfnamefont {M.~T.}\ \bibnamefont
  {Moneck}}, \bibinfo {author} {\bibfnamefont {E.}~\bibnamefont {Yang}}, \ and\
  \bibinfo {author} {\bibfnamefont {J.-G.}\ \bibnamefont {Zhu}},\ }\href
  {\doibase 10.1063/1.4746426} {\bibfield  {journal} {\bibinfo  {journal}
  {Applied Physics Letters}\ }\textbf {\bibinfo {volume} {101}},\ \bibinfo
  {pages} {072411} (\bibinfo {year} {2012})},\ \Eprint
  {http://arxiv.org/abs/https://doi.org/10.1063/1.4746426}
  {https://doi.org/10.1063/1.4746426} \BibitemShut {NoStop}%
\bibitem [{\citenamefont {Getzlaff}(2007)}]{getzlaff2007fundamentals}%
  \BibitemOpen
  \bibfield  {author} {\bibinfo {author} {\bibfnamefont {M.}~\bibnamefont
  {Getzlaff}},\ }\href@noop {} {\emph {\bibinfo {title} {Fundamentals of
  magnetism}}}\ (\bibinfo  {publisher} {Springer Science \& Business Media},\
  \bibinfo {year} {2007})\BibitemShut {NoStop}%
\end{thebibliography}%
	
\end{document}